\documentclass[12pt]{article}
\usepackage{graphicx,amsmath,amsfonts,amssymb,boxedminipage}

\begin{document}
\newcommand{\fig} [1] {Figure~(\ref{#1})}
\newcommand{\mybox} [1] {Box~(\ref{#1})}
\newcommand{\figs}[2] {Figures~(\ref{#1} and \ref{#2})\,}
\newcommand{\eqn}[1] {Equation~(\ref{#1})}
\newcommand{\bs}{\mathbf{S}}
\newcommand{\bsig}{\boldsymbol{\sigma}}
\newcommand{\bdelta}{\boldsymbol{\delta}}

\baselineskip 24pt

\begin{center}
{\huge Spike sorting in the frequency domain with overlap detection}\\
\vspace{1 cm}
\end{center}

{\bf \noindent  
Dima Rinberg$^{1}$\\
William Bialek$^{2}$\\
Hanan Davidowitz$^{3}$\\
Naftali Tishby$^{4}$\\}
{\it NEC Research Institute, 4 Independence Way, Princeton, NJ 08540.}\\
\vspace{0.5 cm}

\noindent
{\it Present addresses:\\
$^{1}$ Monell Chemical Senses Center, 3500 Market St. Philadelphia, PA 19104.}\\
$^{2}$ {\it Dept. of Physics, Princeton University, Princeton, NJ
08544.}\\
$^{3}$ {\it  PharmaSeq, Inc., 1 Deer Park Dr., Suite F, Monmouth Junction, NJ 08852.}\\
$^{4}$ {\it Institute of Computer Science and Center for
Neural Computation,} \\
$^{\;}$ {\it The Hebrew University, Jerusalem 91904, Israel.}
\vspace{0.5 cm}

\pagebreak

\begin{abstract}
\baselineskip 24pt
This paper deals with the problem of extracting the activity of
individual neurons from multi-electrode recordings. Important aspects
of this work are: 1) the sorting is done in two stages - a statistical
model of the spikes from different cells is built and only then are
occurrences of these spikes in the data detected by scanning through
the original data, 2) the spike sorting is done in the frequency domain, 3)
strict statistical tests are applied to determine if and how a spike should
be classiffed, 4) the statistical model for detecting overlaping spike
events is proposed, 5) slow dynamics of spike shapes are tracked during long
experiments. Results from the application of these techniques to data
collected from the escape response system of the American cockroach,
Periplaneta americana, are presented.
\end{abstract}

\begin{center}{\bf Keywords}\end{center}
\baselineskip 24pt
extracellular recording, clustering, overlap detection,
multicellular, gaussian, waveform variability, cockroach

\pagebreak
\baselineskip 24pt

\section {Introduction}
Classical methods for exploring the mechanisms of brain function
involve recording the electrical activity of single nerve cells.  Much
has been learned from this approach, but there are several
reasons to go beyond single neuron recording \cite{rieke}.
First, multineuron
recording greatly increases the efficiency of studying
the properties of single neurons.
Second, recording simultaneously from many
neurons allows access to the
precise temporal relations among action potentials in multiple
neurons \cite{usrey}.
This can provide a testing ground for the hypothesis that these temporal
relationships carry
significant information.
Finally, multineuron recording experiments might give us a glimpse into the
collective dynamics in neural networks, {\em e.g.}
the existence of multiple stable states and possibility of switching between
them
 \cite{amit, abeles2}.

This paper is concerned with one of the many
technical problems that arise in trying to
adddress the above questions, namely the problem of
sorting out signals from multiple neurons as they
appear on multiple electrodes
 \cite{mcnaughton,reece}.
In a multineuron recording each cell can appear on
multiple electrodes and multiple cells can appear on
each electrode.  
Finding the spike times for each cell is difficult because
spikes from different cells are very similar in shape,
making it hard to distinguish among them.
In addition, the events which might be most interesting, synchronous spiking
of nearby cells, are among the most difficult to
disentangle.

\section{The spike sorting problem}
Throughout the nervous system, cells generate stereotyped electrical
pulses termed action potentials or spikes.  
Thus, in the absence in the absence of noise or variability,  
it is expected that each neuron
in the system would generate a characteristic signal at each electrode.
This waveform can be viewed as a single point in signal space.
In the presence of noise, however, these discrete points spread
into clusters of points, each point representing a single spike.
One of the objectives of the spike sorting method described here
is to make a model of these clusters
that is accurate enough to allow
the assignment of any given voltage waveform to one of the clusters.
Even though it is unlikely that all the details of
the cluster model are correct, our hope is that any such errors
will not hamper spike classification.

If $\cal V$ is the set of voltages measured on all of the electrodes during
a short time segment, the probability distribution of
these voltages may be decomposed as
\begin{equation}
P({\cal V}) = \sum_c P({\cal V}|c) P(c) ,
\label{probv}
\end{equation}
where $P(c)$ is the total probability of observing a spike from
cell $c$ and $P({\cal V}|c)$ is the distribution of voltage waveforms
that arise from this cell.
Conversely, if a set of voltage signals ${\cal V}$ is observed,
the probability that it
comes from cell $c$ can be derived from \eqn{probv} by applying Bayes' rule,
\begin{equation}
P(c| {\cal V}) = {{P({\cal V}|c) P(c)}
\over
{{\sum_{c'} P({\cal V}|c') P(c')}}}.
\label{probcv}
\end{equation}
Ideally one would like assignments which are certain.
Statements of the form ``probably a spike from cell 2, but maybe from cell
1,'' ``probably a spike from cell 3, but maybe overlapping spikes from
cells 4 and 5,'' defeat our purpose.
To avoid this type of problem the
distributions $P({\cal V}|c)$ and $P({\cal V}|c')$ should not overlap when
$c\neq c'$.
Furthermore, it is hoped that if these distributions overlap only slightly,
a precise
model for the distribution of each cluster may not be necessary and
a small number of parameters may suffice to make reliable distinctions
among the clusters.
If a reliable model of this
form can be made, then the assignment problem is solved.
Note also that, in this limit, our prior assumptions
about the likelihood of different neural responses
plays no role.

The approach to the clustering problem described here is as follows.
First it is assumed that individual clusters
$P({\cal V}|c)$ have a Gaussian form in which each frequency
component fluctuates independently.
Next, the best possible clusters are found and the mean
and variances at each frequency are calculated for every cluster.
If this Gaussian model were correct, the probability that a given
cluster could generate a particular waveform would depend only on the
$\chi^2$ distance
between that waveform and the cluster mean, with appropriate weighting by
the
variances.
These  $\chi^2$ values provide a set of new dimensions along which
distributions should be nonoverlapping if certainty in
the assignments has been achieved.
In addition, it is possible to check if the clustering is self consistent
because
every member waveform
of a given cluster will yield a distance of  $\chi^2 \approx 1$ for that
cluster and $\chi^2 >> 1$ for all of the other clusters.

This combination of $\chi^2$ below a threshold for one cluster and above
threshold for all other clusters is the signature of unambiguous assignment.
The reader should note that we achieve this result despite the fact that
the real clusters need not obey the assumptions of our simple model.
This can be seen by looking closely at the form of the resultant $\chi^2$
distributions.
The implementation of these ideas takes on the following form.
First, a statistical model of the spikes from different cells is
built by clustering.
Second, in a completely independent stage we detect the occurrence of these
spikes in the data.
This is initially done for single spikes.
Finally,  by superimposing these statistical models,
overlaping spikes are treated as well.

We emphasize once more that we are {\em\underline{not}} proposing
the Gaussian model as an exact model of the relevant
probability distributions.  On the contrary, the goal is
to show that the Gaussian model suggests dimensions along
which clusters are discriminable, and that the real clusters
are almost perfectly discriminable along these dimensions.
Once this is established, the precise form of the model is
irrelevant.  In cases where nonGaussian behavior is known
to be crucial, one might start with different assumptions
but the general strategy would be the same:  we want to
exhibit explictly the discriminability of clusters along
some small number of relevant directions in waveform space, and
the starting model is just an aid to finding these dimensions.
Similar ideas using a $\chi^2$ test were proposed in recent work by Pouzat et~al. \cite{pouzat}.

\section {The spike sorting algorithm}
The first step is to build the set of clusters.
This is done by identifying the different spike
shapes in the data and constructing a statistical model for each
recognized spike type.
As with many other spike sorting algorithms
the work described here is based on the following assumptions:
a) different spikes from the same cell are very similar,
b) spikes from different cells have different waveforms on at least one
electrode,
c) if a cell fires once, it fires many times and
d) overlaps are fairly rare.

\subsection{Clustering}
\label{sec:clustering}
Before the actual clustering can be done a
large number of each of the different spike types is needed.
To do this the data is broken into short frames.
The content of each frame is then examined to see if it contains ``clean''
spikes
which are described below.
A frame refers to a set of data snippets, one from each
electrode at a given time.
We emphasize that during this initial pass through the data we are interested only in collecting
clean representatives of every spike type and no attempt is made to deal
with frames that did not obviously contain only one single spike.

The object of the clustering is to group similar spikes together.
This must work in spite of the fact that
two similar spikes may be shifted slightly in time  \cite{lewicki}.
To deal with this problem great care is given to accurately aligning the
spikes 
during the clustering process.

\textsf{\textbf{Objects to be clustered-}}
While we measure time domain signals,  from
now on each frame, $f$, is represented by a vector composed of
the concatenated Fourier
transforms of the voltage waveforms of the data from each
of the $N_e$ electrodes, {\em i.e.,}
\begin{equation}
{S}_{f}(\omega)=\left[{S}_{f,1}(\omega),
{S}_{f,2}(\omega),
\cdot\cdot\cdot S_{f,N_e}(\omega)\right].
\label{clusterobjects}
\end{equation}
The alignment of spikes in the frequency domain is achieved simply by
multiplying the Fourier components  by $e^{i\omega\tau}$ where $\tau$ is the necessary time delay (see
below).
It should be noted that although the work presented here deals
with sorting concatenated spectra,
other objects can be sorted as well.
Extensive experiments on sorting different objects were
done while developing the methods described here but
for our data ${S}_{f}(\omega)$ proved to be the most useful.
For example, reference
 \cite{rinberg}
describes the sorting of transfer
functions between electrodes which is independent of the spike shape.
This may be of interest in cases where the spike shape
can change \cite{mitra} ({\em e.g.} in bursting cells).
In certain cases power spectra, which are invariant to time shifts, can be
used as well.

\textsf{\textbf{Description of the algorithm-}}
The clustering algorithm is outlined in Box(1).
Its 
various steps are described below.

\begin{itemize}
\item{\bf Line 1: Initialization-}
First all frames are averaged yielding an
average signal,
\begin{align}
    \bar{S}_{0, e}(\omega) &= \frac{1}{N_f}\sum_{f=1}^{N_f}
    S_{f,e}(\omega)
    \label{nouselabel1}
    \\
    \intertext{and variance,}
    \sigma^2_{0,e}(\omega) &= \frac{1}{N_f}\sum_{f=1}^{N_f}
    \left|S_{f,e}(\omega)-\bar{S}_{c,e}(\omega)\right|^2
    \label{nouselabel2}
\end{align}
where $N_f$ is the number of frames, $\omega$ is the discrete frequency
index and
the indices $f$, $c$, and $e$ refer to the frame, cluster and electrode
respectively.
For this initial averaging the time
shifts, $\tau_f$, described below, are set to zero for all frames.

\item{\bf Loop 2: Split clusters-}
In every pass, except the first, each cluster is split in two using a small
random vector $\delta_{c,e}$, {\em i.e.},
$\bar{S}_{c,e}\rightarrow\bar{S}_{c,e}\pm\bar{\delta}_{c,e}$.
This is repeated until some criterion is met.
Establishing this criterion proved to be
a difficult problem to solve generally.
An a priori knowledge of the
expected number of cells  was found to be the most
reliable criterion for stopping the cluster splitting.

\item{\bf Loop 3: Reassign frames-}
This loop executes an expectation maximization algorithm.
The average and variance at each frequency component of every cluster are
calculated 
using the appropriate time delay for every member frame calculated against
each cluster.
The time delay $\tau_f$ that minimizes $\chi^2_{f,c}(\tau_f)$
is calculated against all clusters.
The frame is then reassigned to the
cluster that matches it most closely, {\em i.e.}, the one that yields the
minimum

$\chi^2_{f,c}(\tau_f)$. This is done until the frames are distributed in
such a way 
that these parameters no longer change \cite{papoulis}.

\item{\bf Line 4: Finalize clusters-}
After the clustering ends some clusters may actually be identical within
statistical error.
These are merged into a single cluster.
Others might clearly contain frames of different types and are split
into two clusters.
Still others might contain frames that clearly do not contain spikes
or perhaps contain very few spikes.
These are discarded.
\end{itemize}

\subsection {Detection}
\label{sec:detection}
The end result of the clustering phase is a statistical template for all of
the spike
types  found in the data.
In the next and final phase the data is scanned to find all occurrences of
each spike type. 
The basic idea of the detection is to cut the data into short
frames and determine which cluster best describes the data in that frame and
its precise timing.
Spikes that are well centered in the frames are detected and subtracted from
the data.
The data is then reframed and the process is
repeated until no single spikes remain in the data.
Finally, this process is repeated for overlaps.
This process is shown schematically in \fig{detfig}.

\textsf{\textbf{Single spike detection-}}
Single spike detection is done by finding the $(c,\tau_f)$ pair that
minimizes
\begin{equation}
\chi^2_f(c,\tau_f) =
\underset{\omega,e}{\sum}{{1}\over{\sigma^2_{c,e}(\omega)}}
\left|{S_{f,e}(\omega)-
\bar{S}_{c,e}(\omega)\cdot e^{i\omega \tau_f}}\right|^2
\label{c2ssd}
\end{equation}
for every frame. 
In practice \eqn{c2ssd} is expanded into three terms,
\begin{align}
\chi^2_f(c, \tau_f) &= A(c)+B(c)+2C(c, \tau_f), \\
A(c) & = \underset{\omega,e}{\sum}{{1}\over{\sigma^2_{c,e}(\omega)}}
{|{\bar{S}}_{c,e}(\omega)|}^2,\\
B(c) & =
\underset{\omega,e}{\sum}{{1}\over{\sigma^2_{c,e}(\omega)}}
{|{S}_{f,e}(\omega)|}^2,\\
C(c,\tau_f) & =
\underset{\omega,e}{\sum}{{1}\over{\sigma^2_{c,e}(\omega)}}
\mathrm{Re}\bigl({{S}_{f,e}(\omega){\bar{S}}^*_{c,e}(\omega)\cdot e^{-i\omega \tau_f}}\bigr).
\end{align}
$A(c)$
can be calculated in advance once the frame clustering has been done.
$B(c,\tau_f)$ and $C(c,\tau_f)$  are
calculated in the detection algorithm,
outlined in pseudo-code in Box (2).
The main ideas are described below.

\begin{itemize}
\item
{\bf Line 1: Frame data-}
The complete data set is broken up into nonoverlapping equal sized frames.
\item
{\bf Loop 2: Process each frame-}
It is assumed that a frame can contain either noise (see next section), a
spike, part of a spike or an
overlap of  2 spikes.
Frames containing noise are discarded leaving only those containing single
or multiple 
spike events.  
\item
{\bf Loop 3: Check fit to each cluster-}
Here the chosen frame is compared to each cluster.
Since the spike in the frame may not be centered it is
necessary to align the frame to the cluster.
For each frame the time delay, $\tau_f$, that maximizes the
cross correlation term, $C(c,\tau_f)$, is found for each cluster.

\item
{\bf Line 4: Calculate
$\mathbf{\boldsymbol{\chi}^2_{f,c}(\boldsymbol{\tau}_f,c)}$ for
all clusters-} 
If the spike is not near the  edge of the frame, {\em i.e.},
$\left|\tau_f\right|<\tau_{th}$, then
$B(c,\tau_f)$ is calculated yielding the final term in \eqn{c2ssd}.
All of the terms of $\chi^2_{f,c}(\tau_f,c)$ have now been calculated
yielding an estimate of
the similarity of the frame to each of the clusters.

\item{\bf Line 5: Finalize spike detection-}
The cluster that yields the smallest value for $\chi^2_i(\tau,c)$ is the
cluster most similar
to the frame being tested.
It is not enough to find the cluster for which $\chi^2$ is smallest.
A fit is accepted only if $\chi^2_i(\tau,c) < \chi^2_{th}$.
Frames that were not good matches to any cluster most likely contained an
overlap, 
noise or a partial spike that will most likely be found when the frames are
shifted in a later pass
through the data.
\end{itemize}

\textsf{\textbf{Multiple spike (overlap) detection-}}
Once all of the single spikes have been detected and removed, overlapping
spikes are 
detected and removed as well.
Here we generalize the single spike case to two spikes,
thus, we look for the set of $(\tau_1,\tau_2,c_1,c_2)$ that minimizes
\begin{equation}
\chi^2(\tau_{f,1},\tau_{f,2},c_1,c_2) =
\underset{\omega,e}{\sum}
{1\over{\sigma^2_{c_1,c_2, e}}}
\left|S_{f, e}(\omega)-S_{c_1, e}(\omega)
\cdot e^{i\omega \tau_{f,1}}-
S_{c_2, e}(\omega)\cdot e^{i\omega \tau_{f,2}}\right|^2.
\end{equation}

The algorithm used to find the clusters and time delays is very similar to
that
used for the single spike case with some differences.
First, the minimization executed on line 3 of Box (2) is over two
variables
$\tau_{f,1}$ and $\tau_{f,2}$.
This is computationally intensive
but since there are a
finite number of possible time delays  it remains manageable on a personal
computer.
As before many of the calculations can be performed once the cluster centers
are known.

Another issue which needs to be addressed when dealing with two spikes is
the calculation of
$\sigma^2_{c_1,c_2,e}(\omega)$.
Here we assume that there is an inherent noise in the spike shape and an
additive background noise.
Thus the variances are given by
\begin{equation}
\sigma^2_{c_1,c_2, e}\left(\omega\right)=
\sigma^2_{c_1,e}\left(\omega\right)+
\sigma^2_{c_2,e}\left(\omega\right)-
\sigma^2_{n}\left(\omega\right),
\end{equation}
where $\sigma^2_n(\omega)$ is the background noise computed from regions
devoid of spikes.
While this assumption is not true in general it is reasonable
since the variances of the different clusters are
assumed to be independent.
It proved to be a reliable working model.
This measure of the two-spike variance takes into account the contribution
of
each cluster  but does not overcount the background noise.

\section {Application to multi-electrode data}
The techniques
described above were applied to data recorded from the escape response
system 
of the American cockroach {\em Periplaneta americana}
\cite{camhi1, kolton, camhi2, westin, camhi3}.
Neural activity was recorded using 8 hook electrodes attached to the two
bilateral
abdominal connectives.
In this arrangment each electrode measures a weighted sum of
the activity from the different neurons in the connective.
A more detailed description of the experimental setup is given in
reference \cite{rinberg2}.
Typical experiments lasted for several hours and yielded about 8 GB
of raw data (see \fig{rawdata}).
The aim of
the work described here is to unravel the individual
spike times from this multi-electrode data.

\subsection{Statistical model of the clusters}
Here we describe the application of the clustering algorithm described in
Section (\ref{sec:clustering}) to the multielectrode cockroach data.

\textsf{\textbf{Frame selection-}}
Only frames that fulfill the following criteria were collected to produce a
model of the clusters:
a) the signal in the middle
of the frame is above some threshold, $v_{th}^{mid}$, on at least one
electrode and 
b) the signal at the frame edges is less than some other threshold,
$v^{edge}_{th}$, on all of the
electrodes. 
This idea is illustrated in \fig{framing}.

The thresholds were proportional to the background noise levels.
This background signal, $\bar{v}_b$, was the average
of several averages,
\begin{equation}
v_b = (\frac{1}{N_p}\sum_{i=1}^{N_p}\left|v_i-\bar{v}\right|^2)^\frac{1}{2},
\end{equation}
calculated from
regions devoid of spikes at the beginning,
middle and end of the experiment (usually silences between trials when no
stimulus was presented).
Here $v_i$ are the data points while the number of data points
was typically about $N_p=5\cdot10^4$.
The thresholds described above were defined as
$v_{th}^{mid}=4 \cdot\bar{v}_b$ and $v_{th}^{edge}=1.5\cdot\bar{v}_b$.
These thresholds worked well for our data but
will likely be different for data from other experiments.

While the choice of thresholds is to some extent arbitrary, there are
several
guidelines that can be followed.
If the threshold is set too high low energy spikes will be missed.
If it is set too low, a large cluster containing noise will appear.
We have checked that lowering the thresholds does not
influence the larger amplitude clusters identified
by our algorithm.  
The setting of the threshold is
thus a compromise between identifying all the small
amplitude spikes and minimizing computing time.

To avoid missing spikes because of voltage drifts or overlapping tails of
nearby spikes, the
frames are first  detrended by subtracting a linear fit to the first and
last 0.5 ms of
each frame.

\textsf{\textbf{Time shift between electrodes-}}
Spikes appear on each electrode at different times because of the finite
propagation
time of the action potentials  along the neurons in the connective.
This is  evident in the data shown in \fig{framing}.
Frames are thus defined with time delays between the different channels.
A further complication arises from the fact that the
time delays on the different electrodes
may be different for different cells.
An average delay was found by  calculating the time of the
maximum of the cross correlation of voltage traces from the different
electrodes.
Because these delays can change during the course of an experiment these
inter-channel delays were in turn averaged from widely separated segments of
data. 
Typically, these delays were between 0.1 and 0.8 ms depending on the
positioning
of the electrodes.

\textsf{\textbf{Frame size-}}
To keep calculation times short, a small frame size is desirable.
On the other hand, the frames have to be big enough to account for the
different propagation times of the different cells.
For the experiments described here frames were 3.2 ms long (64 data
points).

The clustering begins after 10,000 frames containing candidate spikes
have been collected.
While each frame is represented by the vector defined in
\eqn{clusterobjects}
only the low frequency components (the first 16 complex numbers)
of each Fourier transform were used because the
higher frequencies were found to be indistinguishable from noise.
Thus ${S}_{f,e}(\omega)$ was of dimension
$\frac{1}{2}\cdot32\cdot N_e$ complex numbers, where
$N_e$ is the number of electrodes.

Since about 7 cells are expected on each side the clustering is stopped once
16 clusters have been found on a side.
Up until this point the clustering is automatic.
Some results of this automatic clustering are shown in \fig{autosort}.
Intervention is now needed to refine the clustering.
To do this the $\chi^2$ distributions of the clusters are examined manually.
Clusters are then merged, split or discarded as necessary.
This process could be automated, but was not deemed worth the effort.
Results of the clustering after the final manual intervention are
shown in \fig{mansort} and \fig{dotplot}.

\textsf{\textbf{Track slow changes-}}
Thus far,  the first $\approx10^4$ frames have been used to establish the
statistics of each cluster.
The accuracy of the spike detection
can be improved if the change in spike shapes are tracked during the course
of a long
experiment.
For the cockroach experiments described here, the trial period of 100 s
was chosen as the time scale upon which changes are tracked.
To do this the cluster statistics $\bar{S}_{c,e}(\omega)$ and
$\sigma_{c,e}(\omega)$ are
updated with consecutive 100 s segments of data on a first-in, first-out
basis.
Once the cluster statistics have been computed, the next $N$ ``clean''
frames from
the data are found and are each assigned to the cluster that is closest
based on the $\chi^2$ distance defined in Box (1).
These frames are then added to these clusters
while the same number of the earliest frames
are removed.  
As usual, frames that exceed a certain $\chi^2$ for all clusters are
discarded.
The cluster statistics are then recalculated,
yielding dynamic clusters of time dependent spike statistics
$\bar{S}_{c,e}(\omega, t)$ and $\sigma_{c,e}(\omega, t)$ that
are used as the templates for the spike detection
described in the next section.
Results of spike shape tracking are shown in \fig{tracking}.

\subsection {Detection}
For the detection phase the complete data set is broken into
nonoverlapping frames containing 64 data
points  (3.2 ms).
In practice, $B(c)$ is calculated only after zero padding the
frame on the outside 2 ms.
This is done to eliminate the possibility of a nearby spike, with an
overlapping
tail,  causing  $\chi^2_{f,c}(\tau_f,c)$ to be too large.
$C(c,\tau_f)$ does not need to be recalculated
because the clusters themselves are averages of many clean spikes and
therefore 
inherently zero padded.

A fit is accepted only if $\chi^2_i(\tau,c) < \chi^2_{th}$ which
was determined by the shape of the distribution.
Typically $ \chi^2_{th}\approx2$, though this threshold would likely be
different for other data. The probability of detecting a spike is smaller if
it overlaps with the
tails of nearby spikes.
This problem can be greatly reduced if, once spikes have been classified,
the average spike from the corresponding cluster is subtracted from the data
(see \fig{detfig}).
Every pass through the data leaves a smaller number of unclassified
spikes.
After each pass the data is reframed with a time shift of $\frac{1}{4}$
frame and the 
process is repeated.
Ideally, after 4 passes all single spike events have been detected.

\textsf{\textbf{Multiple spike (overlap)detection-}}
Once all of the single spikes have been detected and removed, overlapping
spikes are 
detected and removed as well.
Figures (\ref{singleoverlap} and \ref{doubleoverlap}) show the results of
describing an overlap
as single and double spike events.
Notice that the automatic recognition of an event
as being a two spike event is dependent on being able
to reject the ``best" single spike description of the
event.  
Thus, in \fig{singleoverlap}, the best single spike description
is quite good, and the identification of a spike from cell
(b) is correct, but the value of $\chi^2 = 3.5$ is
unacceptably high, as can be seen from the $\chi^2$
distributions.  
Once we explore the space
of two spike events we find a unique description with $\chi^2 \approx1$
(see \fig{doubleoverlap}).

\section {Discussion}
The spike sorting method described here has several key features.
First, the experimental design is such
that the full waveforms from all electrodes are recorded
continuously during the course of an experiment.
This is an advantage over many
spike sorting techniques that are based on the
idea of feature clustering (see  \cite{lewicki} for a recent review of
spike sorting techniques as well as an  older review in  \cite{schmidt}).
In feature clustering one or more features of the spikes (spike
height, peak to peak amplitude, rise times, etc.) are extracted and
clustered.  
Since only a few features of the spikes are used, subtle differences between
spikes from 
different cells can be lost.
In addition, it is not a priori clear which, and how many,
features to use---are there better (optimal) ones?
When applied properly this technique can work well for a small number of
cells
but does not work well with many cells.

Another important feature of the present technique is that the
construction of models for the spike shapes from
individual cells (clustering) is separated from the problem of finding those
spikes in the data
(detection).
The advantage here is that one has the leisure to look for clean examples of
each spike type and thus to build a good statistical model of each spike
type.
Only then does one look for the occurrence of these spikes in the data.
These models are time dependent in the sense that they track the change in
spike shape
during the course of an experiment.
This has traditionally been a problem in
template-matching spike sorting techniques in which a model of the spike
shapes are
constructed \cite{bergman}.
In this technique, these models are compared to a given spike and a decision
is made 
as to whether it belongs to the class defined by this template or not.
One of the more advanced implementations of this technique \cite{lewicki2}
works well in many cases but relies heavily on a Gaussian model of the
noise. 
Another technique \cite{fee} that makes no assuptions about the waveform
noise uses the spike shapes as well as refractory period statistics to
classify cells. 
It works well with bursting cells but in its present form this
technique does not treat overlaps well.
This is another advantage of the separation of clustering from detection: it
simplifies 
the problem of overlap detection which has traditionally been one of the
most difficult parts of the spike sorting problem.

Many comparisons and computations that go into the sorting process are
much easier when working in
the frequency domain.
The ease in
which temporal alignment is achieved is one of the advantages of
performing the detection in the frequency domain since
sub-bin time shifts in the time domain would require
resampling \cite{lewicki}.
In the frequency domain this is done simply by multiplying the Fourier components by $e^{i\omega\tau}$
which is equivalent to a time shift of $\tau$ in the time domain.
In addition, it is even possible to sort
spikes from bursting cells (in which spike shapes can change drastically
over short time scales)
by ignoring the spike shapes entirely and sorting on transfer function
ratios \cite{rinberg}.
Since this is an independent type of information
it can be used to check the validity of sorting results.
Sorting on the full Fourier
transforms of the voltage waveforms yielded excellent results for our data
but other investigators will likely find other combinations that work better
for 
their data. 

The statistical methods used in this sorting program allow us to decide
whether the clustered spikes really are discriminable.
In addition, this statistical analysis allows us to develop more rigorous
criteria for accepting
or rejecting the possible detection of a spike.

Finally we note that in a typical experiment $\approx400,000$ spikes are
found. 
Of these ~90\% were found with single spike
detection,  ~8\% were found with overlap detection only
~2\% were events that could not be classified.

\section{Summary}
In this paper the problem of unraveling multielectrode neural data has been
addressed. 
Special attention has been paid to the detection of overlapping
spikes which poses obvious difficulties for any sorting method.
The problem of spike sorting has been separated into two independent parts.
First, a statistical model of all possible spikes found in the data is
constructed. 
Only then are these spikes detected in the data using strict statistical
criteria 
to quantify the quality of this detection.
Overlaps are dealt with after all possible single spike
events have been detected.
These techniques were applied to multielectrode data from the American
cockroach
with good results.

\section{Acknowledgements}
The authors would like to thank N. Brenner and R. de~Ruyter~van Steveninck
for many helpful 
suggestions and discussions.
\pagebreak

\bibliographystyle{apalike}

\newpage

%
\begin{centering}
\setlength{\fboxsep}{5pt}
\setlength{\fboxrule}{1pt}
\begin{boxedminipage}[c]{5.5in}
\setlength{\fboxrule}{0pt}
\vspace{0.05in}\framebox[0.25in][r]{\sf 1} \hspace{0.1in} 
  {\sf  set $\tau_f:=0$ and calculate initial $\bar{S}_{c,e}(\omega)$
  and $\sigma_{c,e}(\omega)$}\\
\vspace{0.05in}\framebox[0.25in][r]{\sf 2} \hspace{0.1in} 
  \textbf{\textsf{\large repeat}} {\sf split clusters}\\ 
\vspace{0.05in}\framebox[0.25in][r]{\sf 3} \hspace{0.3in}
  \textbf{\textsf{\large repeat}} {\sf update frame assignment}\\ 
\hspace{0.5in}\framebox[0.25in][r]{\sf } \hspace{0.5in}
{\sf find average and variance for each cluster using $\tau_f$}
  \begin{minipage}[c]{4in}
    \begin{align*}
      \hspace{1.25in}
      \bar{S}_{c,e}(\omega) &:= \frac{1}{N_c}\underset{f\in c}{\sum}
      S_{f,e}(\omega)\cdot e^{i\omega \tau_f}\\
      \sigma_{c,e}(\omega)  &:= \frac{1}{N_c}\underset{f\in c}{\sum}
      \left|S_{f,e}(\omega)\cdot e^{i\omega
      \tau_f}-\bar{S}_{c,e}(\omega)\right|^2
      \intertext{\framebox[0.25in][r]{\sf } \hspace{0.5in}
      {\sf for every frame, $f$, find $\tau_f$ that minimizes}}
      \chi^2_{f,c}(\tau_f)  &:= 
      \underset{\omega,e}{\sum}{{1}\over{\sigma^2_{e}(\omega)}}
      \left|{S_{f,e}(\omega)\cdot e^{i\omega \tau_f}-
      \bar{S}_{c,e}(\omega)}\right|^2 \hfill\notag
    \end{align*}
  \end{minipage} \notag

\framebox[0.25in][r]{\sf } \hspace{0.5in}
{\sf assign frame to the cluster that yields smallest $\chi^2_{f,c}(\tau_f)$}\\ 
\vspace{0.05in}\framebox[0.25in][r]{\sf } \hspace{0.3in}
  \textbf{\textsf{\large until}} {\sf clusters are stationary}\\
\vspace{0.05in}\framebox[0.25in][r]{\sf } \hspace{0.1in}
  \textbf{\textsf{\large until}} {\sf number of clusters $>$ expected number of cells}\\
\vspace{0.05in}\framebox[0.25in][r]{\sf 4} \hspace{0.1in}
  \sf{finalize clustering by merging, splitting or removing clusters}\\
\vspace{0.05in}\framebox[0.25in][r]{\sf 5} \hspace{0.1in} 
  \textbf{\textsf{\large repeat}} {\sf track slow changes in spike shapes}\\ 
\vspace{0.075in}\framebox[0.25in][r]{\sf } \hspace{0.3in}
  {\sf find spikes in next time chunk of data and update cluster statistics}\\ 
\vspace{0.05in}\framebox[0.25in][r]{\sf } \hspace{0.1in}
  \textbf{\textsf{\large until}} end of data is reached
\end{boxedminipage}
\end{centering}

\noindent Box 1: Pseudo code outline of the frame clustering algorithm.
See text for details.

\newpage
%
%

\begin{centering}
\setlength{\fboxsep}{5pt}
\setlength{\fboxrule}{1pt}
\begin{boxedminipage}[c]{5.5in}
\setlength{\fboxrule}{0pt}
\vspace{0.05in}\framebox[0.25in][r]{\sf 1} \hspace{0.1in} 
  {\sf  Frame whole data set with no overlaps}\\
\vspace{0.05in}\framebox[0.25in][r]{\sf 2} \hspace{0.1in} 
  \textbf{\textsf{\large repeat}} {\sf get next frame, $f$}\\ 
\vspace{0.05in}\framebox[0.25in][r]{\sf } \hspace{0.3in}
  \textbf{\textsf{\large if}} {\sf frame has signal above background} 
  \textbf{\textsf{\large then}}\\ 
\vspace{0.05in}\framebox[0.25in][r]{\sf 3} \hspace{0.4in}
  \textbf{\textsf{\large repeat}} {\sf for each cluster, $c$}\\ 
\hspace{0.5in}\framebox[0.25in][r]{\sf } \hspace{0.6in}
{\sf find time delay, $\tau_f$, that maximizes}

  \begin{minipage}[c]{4in}
    \begin{equation*}
       \hspace{1.25in}
       C(c,\tau_f)=\underset{\omega,e}{\sum}\frac{
                 \mathrm{Re}\bigl({{S}_{f,e}(\omega){\bar{S}}^*_{c,e}(\omega)
       \cdot e^{-i\omega\tau_f}}\bigr)}
       {\sigma^2_{c,e}(\omega)}     
       \label{ssd}
    \end{equation*}
  \end{minipage} \notag
  
\framebox[0.25in][r]{\sf 4} \hspace{0.6in}
\textbf{\textsf{\large if}} 
{\sf $\left|\tau_f\right|<\tau_{th}$ \textbf{\textsf{\large then}}
zero pad frame and calculate $\chi^2_{f}(\tau_f,c)$}\\ 
\vspace{0.05in}\framebox[0.25in][r]{\sf } \hspace{0.4in}
  \textbf{\textsf{\large until}} {\sf all clusters checked}\\
\vspace{0.05in}\framebox[0.25in][r]{\sf 5} \hspace{0.4in}
  {\sf find cluster, $c$, that yields smallest $\chi^2_{f,c}(\tau_f,c)$}\\
\vspace{0.05in}\framebox[0.25in][r]{\sf } \hspace{0.4in}
  \textbf{\textsf{\large if}} {\sf
  $\chi^2_{f}(\tau_f,c)<\chi^2_{th}$ keep this spike and subtract it from data}\\
\vspace{0.05in}\framebox[0.25in][r]{\sf } \hspace{0.1in}  
  \textbf{\textsf{\large until}} {\sf end of data reached}\\
\end{boxedminipage}
\end{centering}

\noindent Box 2: Pseudo code outline of the single spike detection algorithm.
See text for details.

\newpage
{\Large Figure captions}

%
%
\begin{figure}[h]
\caption{A schematic outline of the spike detection algorithm.
Spike events are labeled {\sf a...f} and are shown for only one electrode
and only two 
cells for the sake of clarity.
The detection procedure progresses from top to bottom yielding the spike
times of two cells 
shown in bottom of the diagram.
The rectangular boxes represent the frames.
Note that a spike near a frame edge is not detected until the
framing has  shifted enough to more or less center it.
Overlaps are not detected until all of the single spikes have been removed.
In the work described here, both the single spike detection and
the overlap detection processes consisted of 4 passes each.
See text for more details.}
\label{detfig}
\end{figure}  

%
%
\begin{figure}[h]
\centering
\caption{Sample of the
raw data used in the spike sorting.
Typically, signals from 4 electrodes were recorded on
each side of the abdominal connective along with 2 stimulus channels (not
shown).
Three channels from the right connective are shown.
Data was recorded at 20 kS/s, 16 bits per channel.}
\label{rawdata}
\end{figure} 

%
%
\begin{figure}[h]
\centering
\caption{Example showing the framing of the data from three electrodes
(1-3).  
Grey boxes show frames that have been found to contain a candidate spike.
10,000  such frames are collected to generate a statistical model of the
spikes.
Each white box contains a frame that has been rejected for one or more of
the following
reasons: 
a) they contain noise,
b) they contain an overlap or
c) their energy is too high at the frame edges.
Note that in this data there is a time shift in the appearance of the same
spike on the
different electrodes.
This is a consequence of the spike propagation velocity but is not a
necessary  condition for the spike sorting techniques described here.
}
\label{framing}
\end{figure}

%
%
\begin{figure}[h]
\centering
\caption{Results of the automatic frequency domain spike sorting showing
four possible situations.
Each row corresponds to one cluster.
The center columns (labeled 1-4) show the spike shapes as they appear on
different electrodes.
Each trace shows 30 randomly chosen spikes from each cluster.
The left column shows the average spectra of the signal from electrode 2
with
the corresponding variances.
$N_{sp}$ is the number of spikes found in each cluster.
The right column shows the distributions of the $\chi^2$ distance of all
spikes
to this cluster center.
The shaded area (thin lines) shows the distributions for member (nonmember)
spikes.
Row 1 shows a good cluster clearly separated from the others.
Rows 2 and 3 show clusters that overlap each other and should be merged.
Row 4 shows a small cluster with no clear center which will be deleted.
Its members will be reassigned to other clusters.}
\label{autosort}
\end{figure}

%
%
\begin{figure}[h]
\centering
\caption{Resulting clusters after manual intervention.
Some clusters from  \fig{autosort} have been joined, others have been
deleted and 
still others have been split, depending on the $\chi^2$ distribution
resulting from the 
automatic sort.
Note the clear separation of the $\chi^2$ distributions.}
\label{mansort}
\end{figure}

%
%
\begin{figure}[h]
\centering
\caption{Cluster distribution at one frequency component in the complex
plane.
The circles correspond to a distance of $2\sigma$ from the center.
Even though the clouds partially overlap for these components, they are well
separated in multidimensional space.}
\label{dotplot}
\end{figure}

%
%
\begin{figure}[h]
\centering
\caption{The evolution of the spike shape over the course of
an experiment,
shown for a single cluster. }
\label{tracking}
\end{figure}

%
%
\begin{figure}[h]
\caption{Attempt to describe an overlap of two spikes as a single spike
event.
The dotted traces at the top are the actual raw data of an overlap recorded
on 4 electrodes.
The middle traces show four clusters a-d.
The bottom traces show $\chi^2$ calculated for different time shifts.
The minimum $\chi^2$ is 3.5 found by fitting cell b with a time shift, $\tau
\approx 0$ ms.
The thin line traces at the top of the figure show this fit.}
\label{singleoverlap}
\end{figure}

%
%
\begin{figure}[h]
\caption{Same as \fig{singleoverlap} except that a fit is attempted using
two cells.
$\chi^2$ is now a function of two time shifts, $\tau_1$ and $\tau_2$.
The panels on the bottom show $\chi^2$ as a function of these two time
shifts.
The circles are centered at the minimum of $\chi^2$ for each combination of
cells. 
The smallest $\chi^2$ was found for cells b and d.
This fit is plotted as a thin line at the top of the figure.
Note that this fit is considerably
better that the best fit of any single spike event.}
\label{doubleoverlap}
\end{figure}

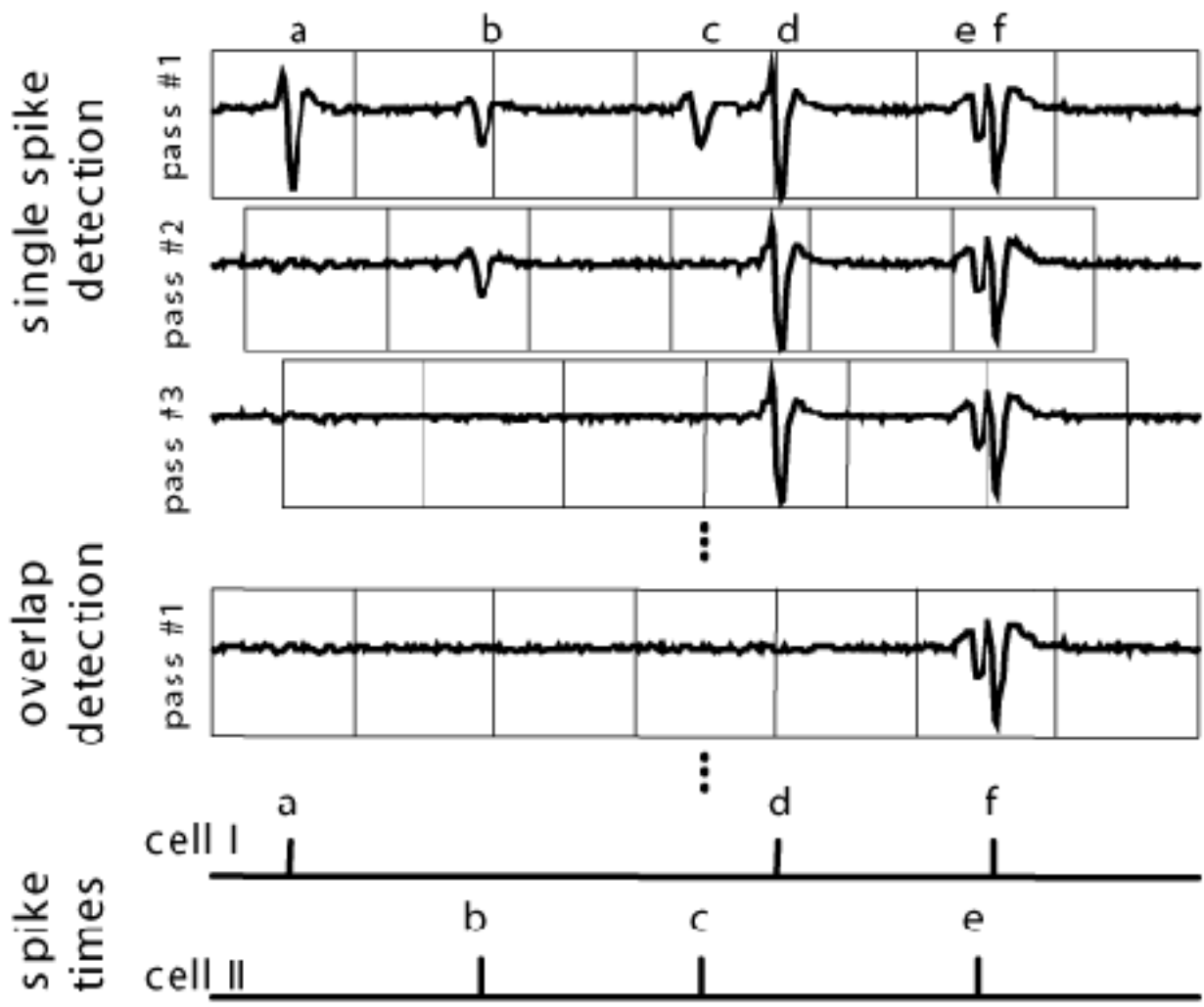
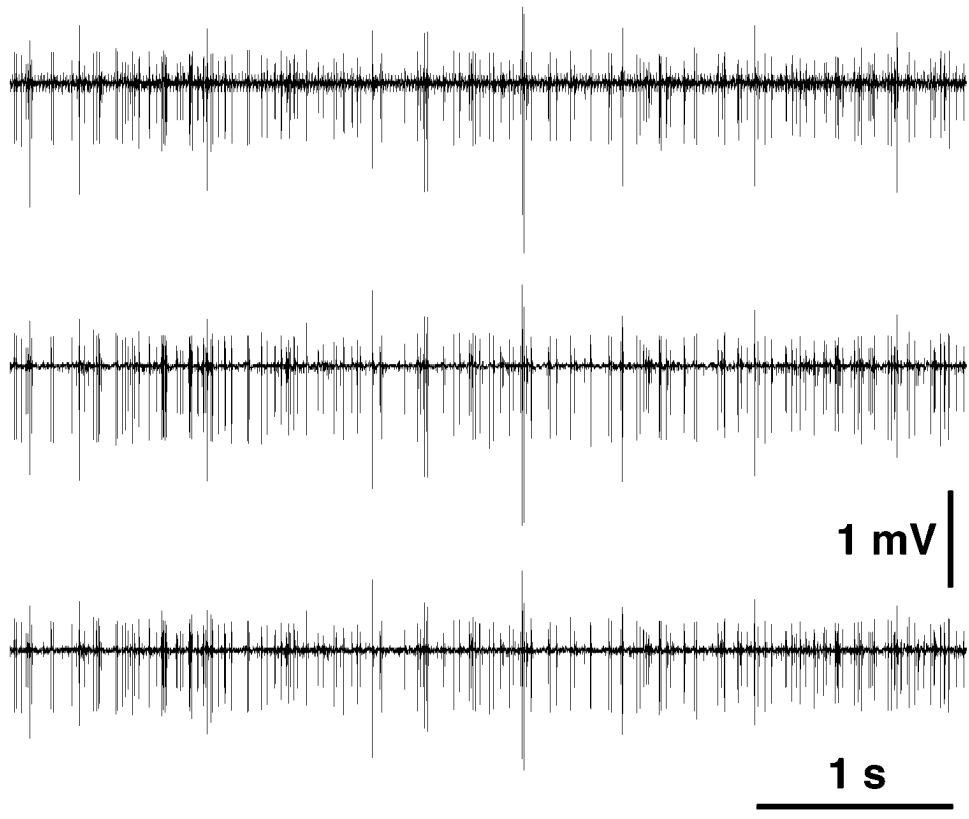
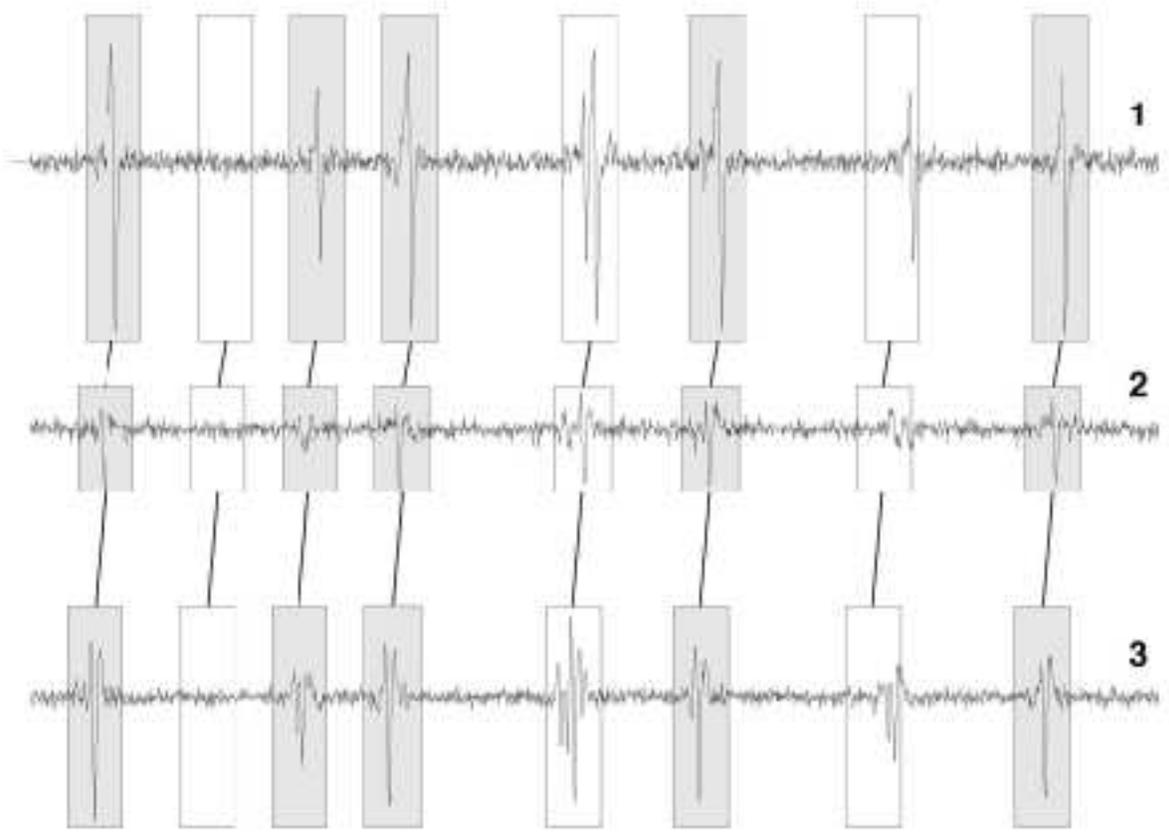
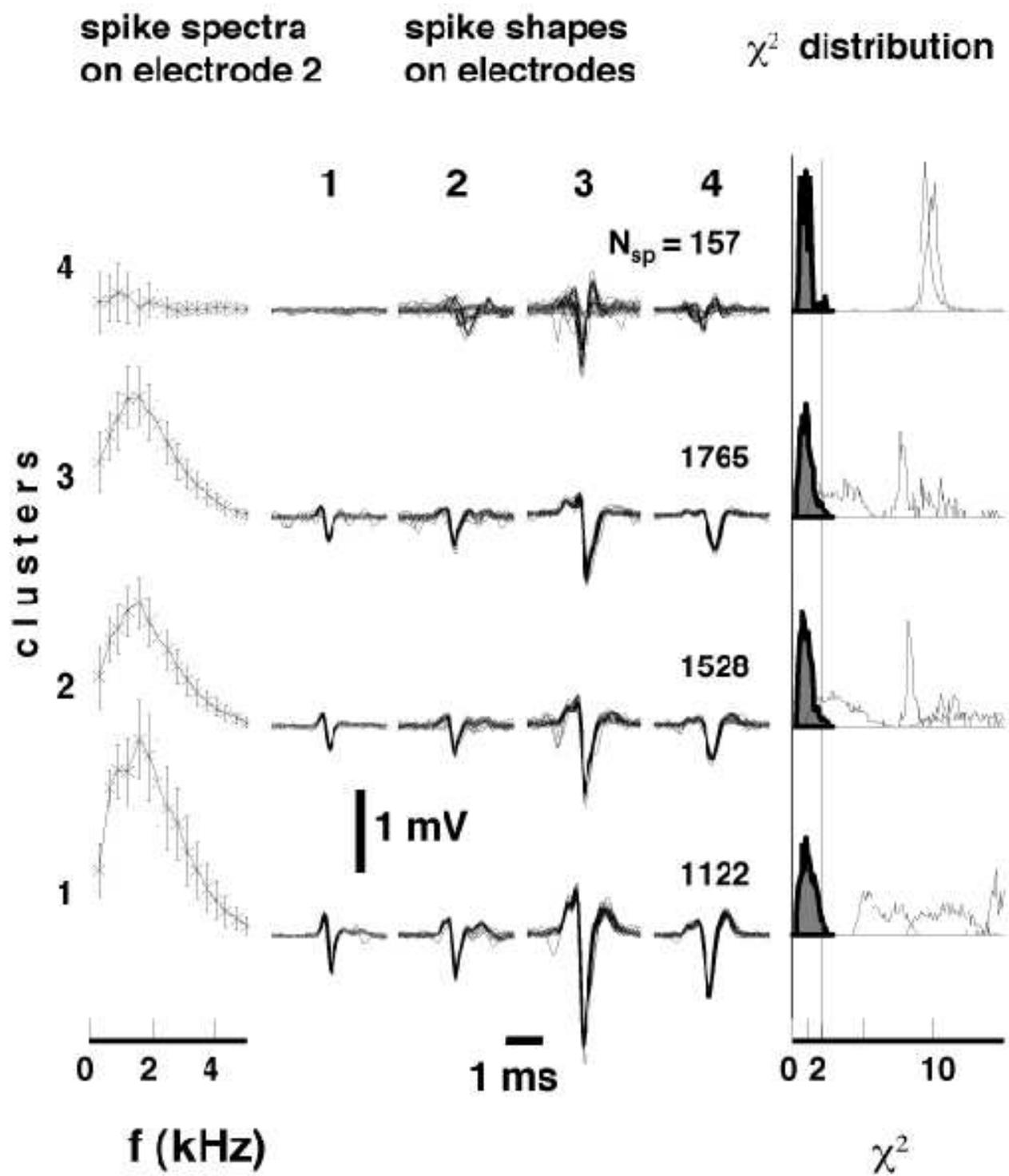
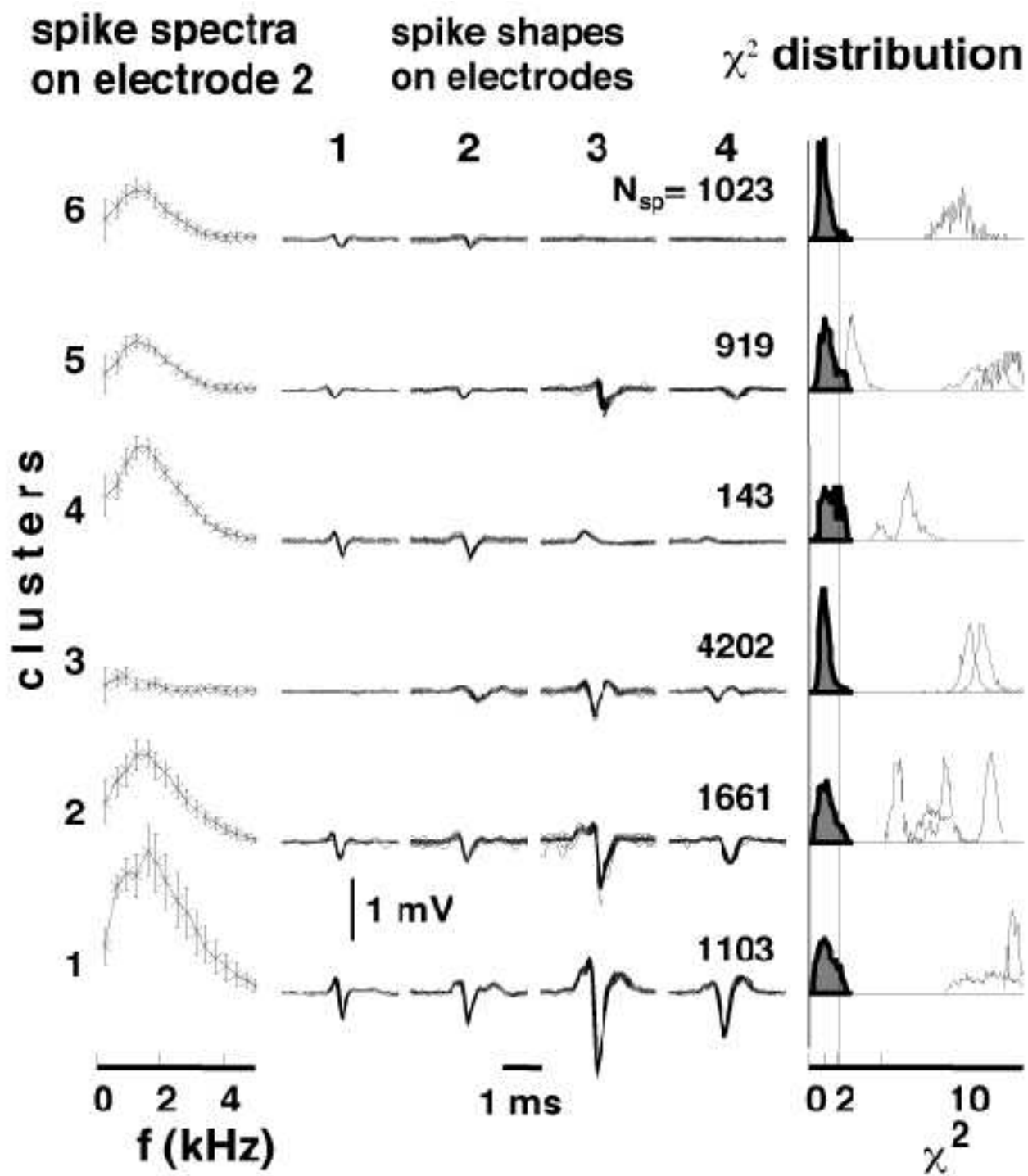
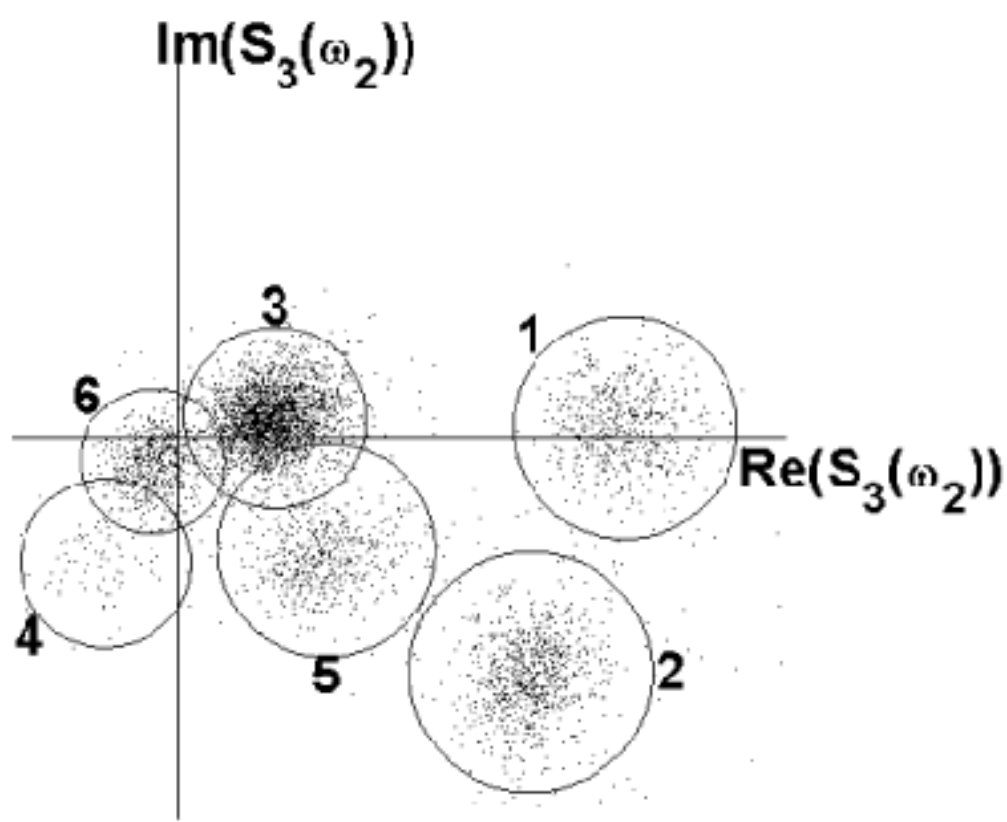
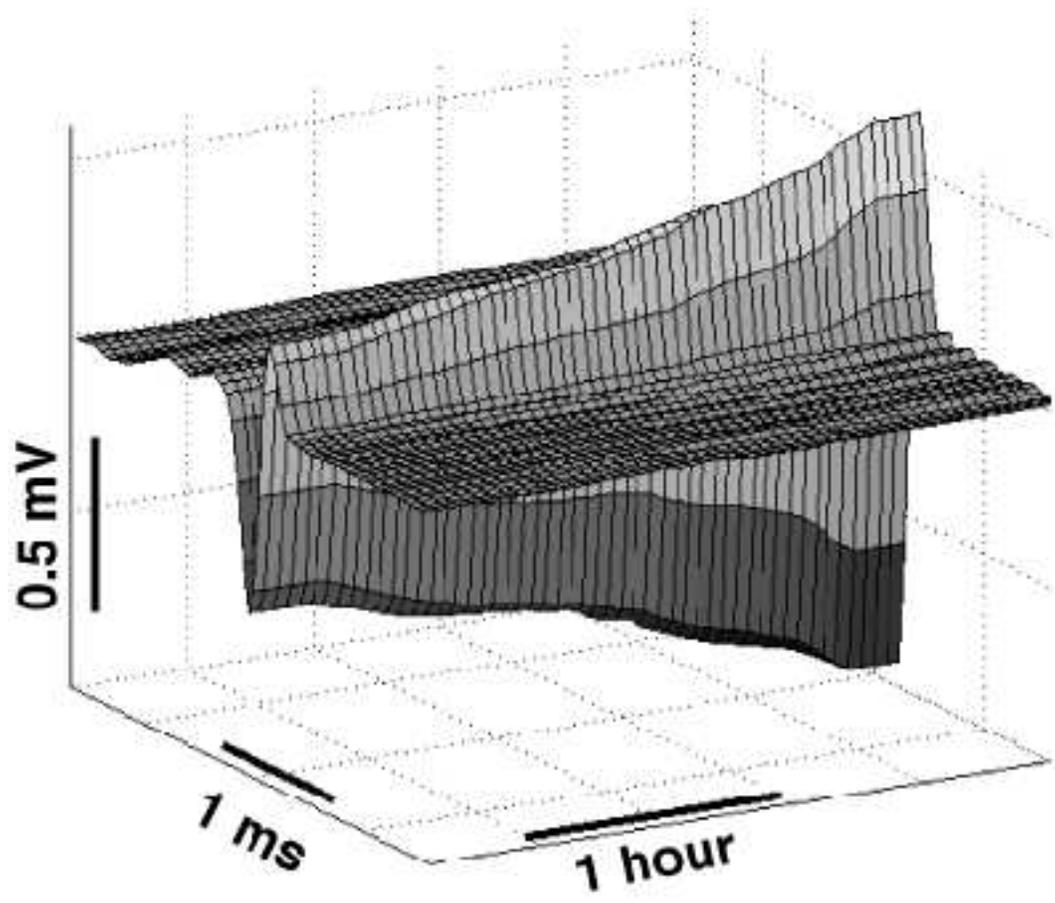
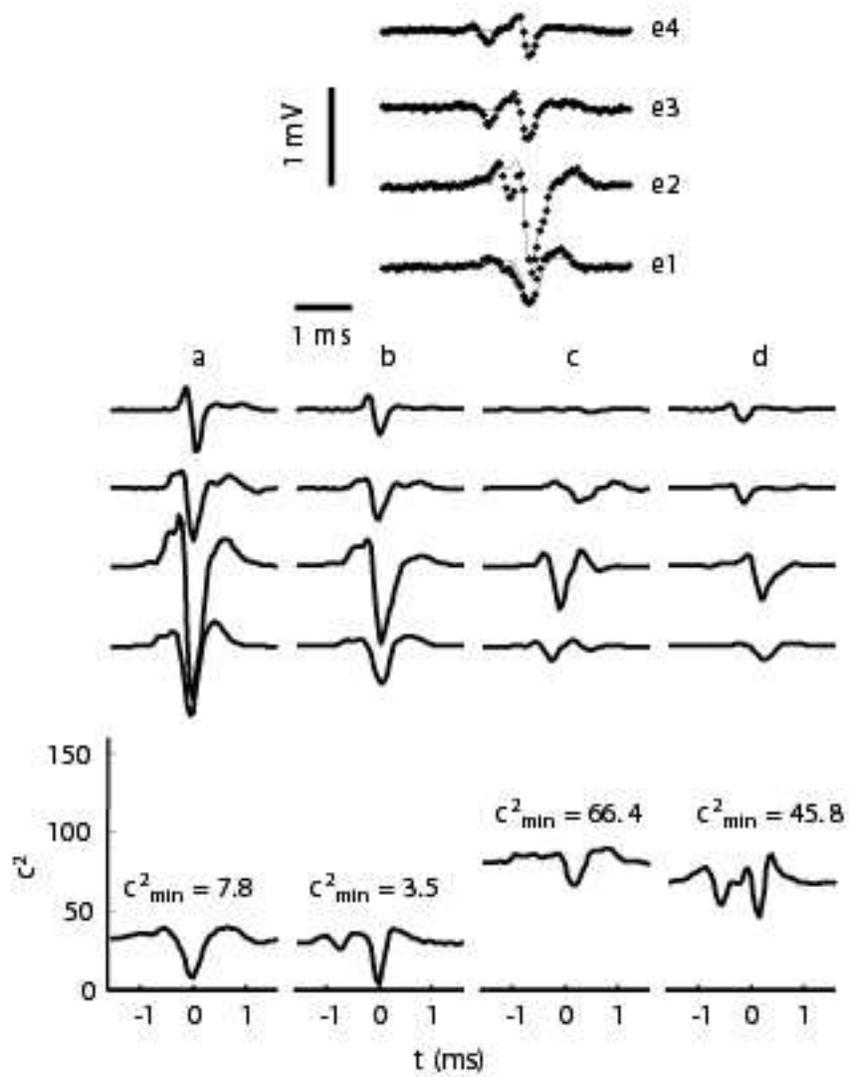
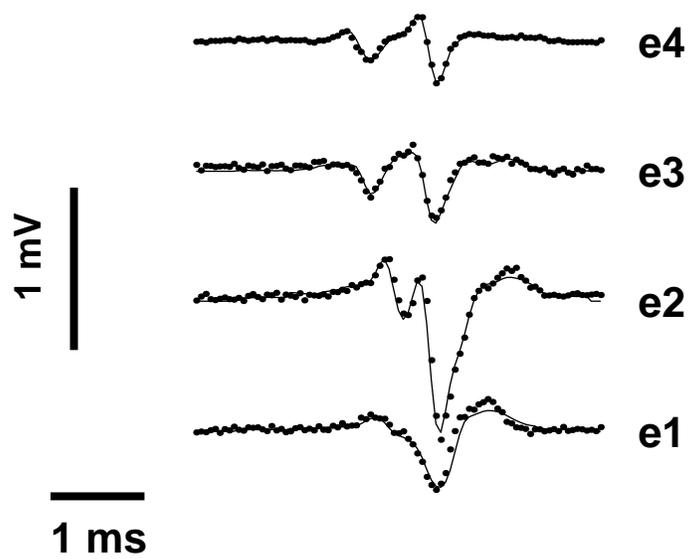
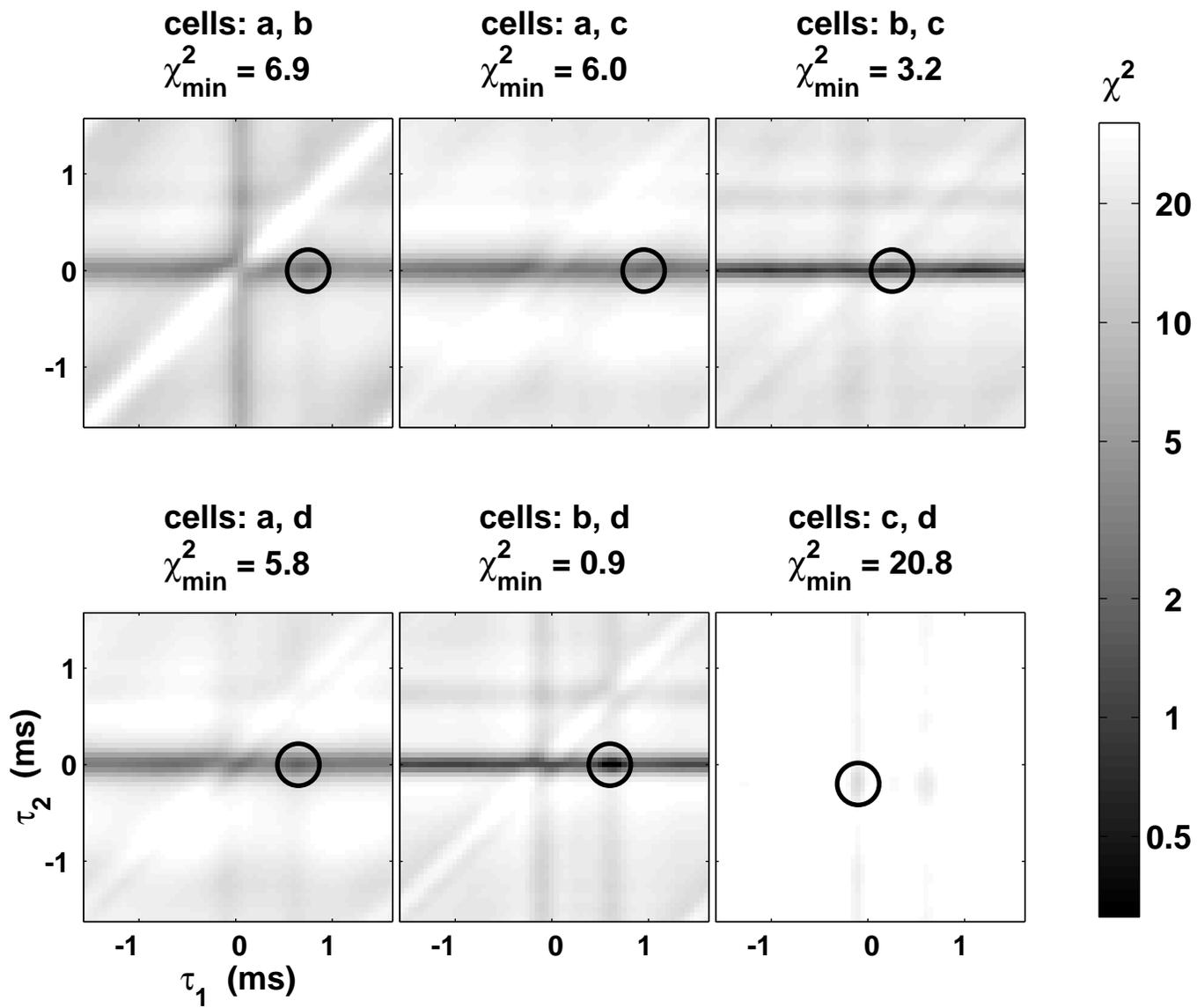
\end{document}